# Hyperspectral Anomaly Change Detection Based on Auto-encoder

Meiqi Hu, Chen Wu, *Member, IEEE*, Liangpei Zhang, *Fellow, IEEE*, and Bo Du, *Senior Member, IEEE*

*Abstract*—With the hyperspectral imaging technology, hyperspectral data provides abundant spectral information and plays a more important role in geological survey, vegetation analysis and military reconnaissance. Different from normal change detection, hyperspectral anomaly change detection (HACD) helps to find those small but important anomaly changes between multi-temporal hyperspectral images (HSI). In previous works, most classical methods use linear regression to establish the mapping relationship between two HSIs and then detect the anomalies from the residual image. However, the real spectral differences between multi-temporal HSIs are likely to be quite complex and of nonlinearity, leading to the limited performance of these linear predictors. In this paper, we propose an original HACD algorithm based on auto-encoder (ACDA) to give a nonlinear solution. The proposed ACDA can construct an effective predictor model when facing complex imaging conditions. In the ACDA model, two systematic auto-encoder (AE) networks are deployed to construct two predictors from two directions. The predictor is used to model the spectral variation of the background to obtain the predicted image under another imaging condition. Then mean square error (MSE) between the predictive image and corresponding expected image is computed to obtain the loss map, where the spectral differences of the unchanged pixels are highly suppressed and anomaly changes are highlighted. Ultimately, we take the minimum of the two loss maps of two directions as the final anomaly change intensity map. The experiments results on public "Viareggio 2013" datasets demonstrate the efficiency and superiority over traditional methods.

## I. Introduction

CHANGE detection for remote sensing images refers to acquiring the difference information of landscapes in the same location by observing it at different time [1-3]. With high spectral resolution, hyperspectral image could distinguish various objects more accurately. Nowadays, HSI change detection has been extensively applied in land use and land cover (LULC) change analysis [4], resource exploration [5], vegetation change analysis [6], and damage assessment [7].

Compared to the conventional hyperspectral change detection, hyperspectral anomaly change detection [8-11] focuses on finding those small and rare changes, which have different distribution with the background changes. The background refers to all ground objects that present at the same location in both HSIs. And the anomalous changes may arise from insertion, disappearance, or movement of small size objects (generally man-made). Besides, the camouflage, concealment and deception of little stationary objects also bring about anomaly changes. And HACD serves as a reminder of focusing on these easily overlooked changes for decision makers [12] and is widely applied in airborne defense and surveillance [13-15].

A straightforward procedure for HACD is to detect anomaly from the difference image of multi-temporal HSIs. However, the spectral variations resulting from diverse imaging conditions in the difference image can result in plenty of pseudo changes, since the solar height angle, illumination and atmosphere condition may have altered tremendously. In theory, these pseudo changes could be suppressed if the multi-temporal images were acquired under the same imaging condition and the unchanged landscapes would show quite similar spectral features. Therefore, anomaly change detection could be done by establishing the mapping relationship from one image to another in order to get a predictive image and then comparing the spectral differences. The idea of predicting the image under the imaging condition of another image is concluded as predictor model [16]. Chronochrome (CC) [17] is such a classical predictor that models the spectral differences of background by the least square linear regression. CC is the first global linear predictor that obtains a predicted image and detects the anomaly changes from the residual image. Since mis-registration error quite influences the performance of CC, the emergence of covariance equalization (CE) [18] solves this problem. CE functions as whitening [19] and is assumed to be able to unify the distribution of two HSIs. Concretely, the two classical methods for HACD, CC and CE, employ the statistical features of the two images to establish the relationship mapping from one imaging space to another. But both of them obey the linear space-invariant assumption, which refers to the affine transformation containing no targets changes for global linear predictors. M. J Carlotto [20] combined clustering with Reed-Xiaoli (RX) [21, 22] to detect man-made changes, which firstly segmented the reference image into several clusters and then used RX to detect anomaly changes within each cluster. RX aims at finding out those pixels that deviate from the main distribution. And the clustering provides additional freedom to adapt to the space-variant background. Eismann et al. [9,16] proposed a new method which divided the background into

M. Hu, C. Wu and L. Zhang are with the State Key Laboratory of Information Engineering in Surveying, Mapping and Remote Sensing, Wuhan University, Wuhan 430072, China (e-mail: meiqi.hu@whu.edu.cn; chen.wu@whu.edu.cn; zlp62@whu.edu.cn).
B. Du is with the School of Computer Science, Wuhan University, Wuhan 430072, China (email: gunspace@163.com).

several classes and mapped one by one to get a prediction. The approach really obtains better detection effect with the segmented linear prediction. But the number of clusters is hard to determine and needs be selected by trial-and-error in the experiment. It is worth noting that most previous predictor models are linear models.

To sum up, the traditional methods use the statistical features of images to construct linear predictors, which suffer from inability of representing the complex relationship of the imaging conditions. Specifically, there are abundant bands of hyperspectral images, which cover a wealth of surface features. Thus a predictor model for HACD requires strong capability of feature extraction. And a nonlinear predictor based on deep learning may give a solution.

Deep learning is composed of multiple processing layers which is similar to neurons of human brain, and is able to learn multiple levels representations of data [23,24]. And deep learning has gained remarkable performance in image processing of remote sensing [25], such as classification [26,27], target detection [28,29], change detection [30,31]. Among deep learning architectures, auto-encoder has shown powerful feature extraction ability in [32], which makes it possible to establish a nonlinear complex mapping relationship. The under-complete auto-encoder (hereafter referred to as AE) is characterized by the design of the bottleneck, which limits the dimension of information transmitted after input layer and helps to extract the crucial features.

Therefore, inspired by the ability of the feature extraction of AE, we proposed a method which employs AE as a nonlinear predictor for HACD called hyperspectral anomaly change detection based on auto-encoder (ACDA). In the proposed method, we utilize two AE networks to get two predictive images, respectively. The special structure "bottleneck" of AE is able to extract the essential features at a lower dimension and reconstruct the input. Using pre-detected unchanged pixels as training samples, the input is the spectral vector of one HSI and the label is corresponding one of another HSI. These two spectral vectors both belonging to the background own consistent essential features but differ in the spectral values. The spectral differences between them result from the different imaging conditions. Thus, the AE network transforms the HSI into another imaging condition to get a predictive image. Then the loss maps will be calculated by computing the mean square error between the predictive images and corresponding expected images. Finally, we take the minimum of the two loss maps as the anomaly intensity map. In addition, the ability of feature extraction and nonlinearity of AE enable the model to deal with the problem of space variant background. The intension of AE predictor is to minimize the spectral differences of multi-temporal HSIs and highlight the anomaly changes. The main contributions of this article are concluded as follows:

1) A nonlinear predictor method based on auto-encoder denoted as ACDA is proposed for hyperspectral anomaly change detection, which gains greater detection performance against other state-of-the-art approaches. And the network structure of the model is compact and simple, which is easy to implement and less time-consuming.

2) By utilizing the bottleneck structure of the AE, the proposed method is effective in extracting the intrinsic information of the high dimensional spectral vector, which is critical to construct the predictive relationship of different imaging conditions.

3) In practice, one of the major issues encountered by HACD lies in the variant background space, which causes the violent and unbalanced spectral variance between two backgrounds of multi-temporal HSIs. Combined the feature extraction and nonlinear mapping, the proposed ACDA obtains good results in this case.

The rest of this paper is organized as follows. Section II gives a representation of the proposed method ACDA. Then we implement the algorithm and two experiments results on real-world datasets are presented in section III. And Section IV draws the conclusion of this paper.

## II. METHODOLOGY

### A. Auto-encoder

Auto-encoder is an unsupervised deep learning network that is able to learn features from the unlabeled data [32]. AE aims at replicating the output from the input and learning the representation of the input, and has broad application, such as dimension reduction [33], image classification [34], and hyperspectral unmixing [35]. Generally, the architecture of AE consists of encoder and decoder, where encoder is used to extract the features of the input, and decoder is designed to decode the feature and reconstruct the input. Concretely, for a single-layer AE network, the encoder is composed of an input layer as well as a hidden layer, and decoder is made up of the hidden layer and an output layer. And joint hidden layer between the encoder and decoder is also called code layer.

There are two significant characteristics of the AE. 1) The neural unit number of input layer equals to the one in the output layer. 2) The size of hidden layer is smaller than the input layer. As for a high-dimensional input, the encoder firstly transforms it into a low-dimensional code, and then the decoder recovers the data from the code. Such a special structure is referred to as bottleneck, which is vital to learn feature representation in unsupervised manner.

For an input sample $z \in \mathbb{R}^P$, the output of the encoder can be written as:

$$E(z) = g\left(w^{(e)}z + b^{(e)}\right) \quad (1)$$

Where $w^{(e)} \in \mathbb{R}^{K \times P}$ is the weight matrix with $K$ features, $b^{(e)} \in \mathbb{R}^{K \times 1}$ is the bias vector, and $g$ refers to the activation function, individually. Then the hidden code is put into the decoder to get a reconstructed result $\hat{z}$:

$$\hat{z} = g\left(w^{(d)} E(z) + b^{(d)}\right) \quad (2)$$

Where $w^{(d)} \in \mathbb{R}^{P \times K}$ is the weight matrix with $P$ features and $b^{(d)} \in \mathbb{R}^{P \times 1}$ is the bias vector, separately. And $\theta$ refers to all parameters $\{w^{(e)}, b^{(e)}, w^{(d)}, b^{(d)}\}$ that need be trained in the

AE network. Given the training sets $z^i, i=1,2,...,N$, the $\theta$ can be iteratively updated by minimizing the reconstruction error which adopts the MSE as the cost function:

$$L(\theta) = \frac{1}{N}\sum_{i=1}^{N}\left(\left\|\hat{z}^i - z^i\right\|^2\right) \quad (3)$$

The AE tries to learn an approximation that the output is as similar as possible to the input. Under the constraint that the unit number of hidden layer is less than the one of the input layer, the code layer is forced to learn a compressed feature representation of the input information. Furthermore, the hidden representation extracts the essential information which can reconstruct the input.

### B. HACD based on auto-encoder

As mentioned above, AE is capable of extracting the essence from the input data due to the bottleneck. Generally, while detecting the anomaly changes from multi-temporal hyperspectral images, we suppose that the ground objects of background do not change, but the spectral features may alter because of the different imaging conditions. Therefore, taken the spectral vector of Time 1 as the training input and the corresponding vector of Time 2 as the training output, the spectral variation of the pair pixel vectors can be fitted by the AE network, since the two spectral vectors of unchanged pixel has the same essential information. AE functions as a predictor that establishes the mapping relationship between two imaging conditions. And anomaly changes could be detected from the residual image where the spectral differences of the background between the predicted image and the expected image are suppressed.

The overview of proposed ACDA is shown in **Fig. 1**. The inputs of ACDA are pairwise spectral vectors of hyperspectral imagery. Then ACDA could be roughly divided into two parts: predictor module and post-processing module. In the predictor module, two systematic AE, whose layers are all fully connected layers, are used to get two predictive images. In **Fig. 1**, the red nodes represent the nodes of input layer, the orange nodes are used to represent the output layer, and the blue nodes are the nodes of hidden layers. After acquiring two predicted images, we choose the MSE between the predictive image and expected image as the loss map. And in the post-processing module, we use the minimum between the two loss maps as the final anomaly change intensity map.

#### 1) AE predictor model

Generally, AE used to reproduce the output from the input. In this article, AE is deployed as a predictor by changing the loss function, where the spectral vector of one HSI is taken as input and the corresponding vector of another multi-temporal HSI is served as the label. We choose the training samples from the pre-detection results of USFA [36,37], which has showed outstanding comprehensive results in [38].

Mathematically, the AE predictor model is defined as follows. Let us denote $X \in \mathbb{R}^{M \times Q}$ as a HSI acquired at Time 1 and $Y \in \mathbb{R}^{M \times Q}$ as another HSI acquired at Time 2, individually, where $M$ and $Q$ refer to the number of pixels and spectral channels, separately. Let us also denote $x = [x_1, x_2, ..., x_Q]^T$ as a training pixel vector of $X$, $y = [y_1, y_2, ..., y_Q]^T$ as the corresponding pixel vector of $Y$.

Supposing there are $n$ hidden layers and $h_i$ denotes the number of nodes of $i$th hidden layer. Given $x$ fed into the AE model, the output of the first hidden layer can be written as:

$$f_1^1(x) = g(w_1^1 x + b_1^1) \quad (4)$$

Where $w_1^1 \in \mathbb{R}^{h_1 \times P}$ and $b_1^1 \in \mathbb{R}^{h_1 \times 1}$ are weight matrix and bias vector, $g$ refers to the activation function. The superscript of $f_1^1$ means the first predictor model mapping from $X$ to $Y$ and subscript of $f_1^1$ means the first hidden layer. The same are true for $w_1^1$ and $b_1^1$. The output of the last hidden layer can be written as:

$$f_n^1(x) = g(w_n^1 f_{n-1}^1(x) + b_n^1) \quad (5)$$

Where $w_n^1 \in \mathbb{R}^{h_n \times P}$ and $b_n^1 \in \mathbb{R}^{h_n \times 1}$ are weight matrix and bias vector, separately. The output of the model can be formulated as:

$$\hat{x} = f(\theta_1, x) = g(w_{n+1}^1 f_n^1(x) + b_{n+1}^1) \quad (6)$$

Where $w_{n+1}^1 \in \mathbb{R}^{Q \times h_n}$ and $b_{n+1}^1 \in \mathbb{R}^{Q \times 1}$ are weight matrix and bias vector, respectively. And $\theta_1$ refers to all parameters $\{w_1^1, b_1^1, ..., w_n^1, b_n^1, w_{n+1}^1, b_{n+1}^1\}$ that need be trained in the first predictor model. Given the training sets $x^i, i=1,2,...,S$ and corresponding label $y^i, i=1,2,...,S$, the loss function of the model can be described as:

$$L(\theta_1) = \frac{1}{S}\sum_{i=1}^{S}\left(\left\|\hat{x}^i - y^i\right\|^2\right) + \lambda_1 \sum_{j=1}^{n+1}\left\|w_j^1\right\|^2 \quad (7)$$

Where the cost function is the MSE, $w_j^1$ refers to weight parameter of different layer and $\lambda_1$ is the regularization coefficient of the first predictor model.

When the predictor has been trained, all pixel vectors of the image $X$ are fed into the predictor to get a predictive image $\hat{X}$. The predictor is not a prophet that predicts what will happen on Time 2, but rather a transformer which maps the landscape from imaging condition of Time 1 to imaging condition of Time 2. Consequently, the imaging condition of the predicted image $\hat{X}$ is the same as the one of expected image $Y$. Furthermore, the spectral differences of the background between $\hat{X}$ and $Y$ are highly suppressed. We use the MSE between the predicted spectral vector and the expected one as the anomaly change intensity. The loss map $I_1$ can be described as the MSE of spectral vector between $\hat{X}$ and $Y$ as follows:

$$I_1 = \text{MSE}(\hat{X}, Y) \quad (8)$$

Considering the change direction of two images, we can get another AE predictor model with predicted image $\hat{y}$ when feeding the pixel spectral vectors of $Y$. The loss function of the

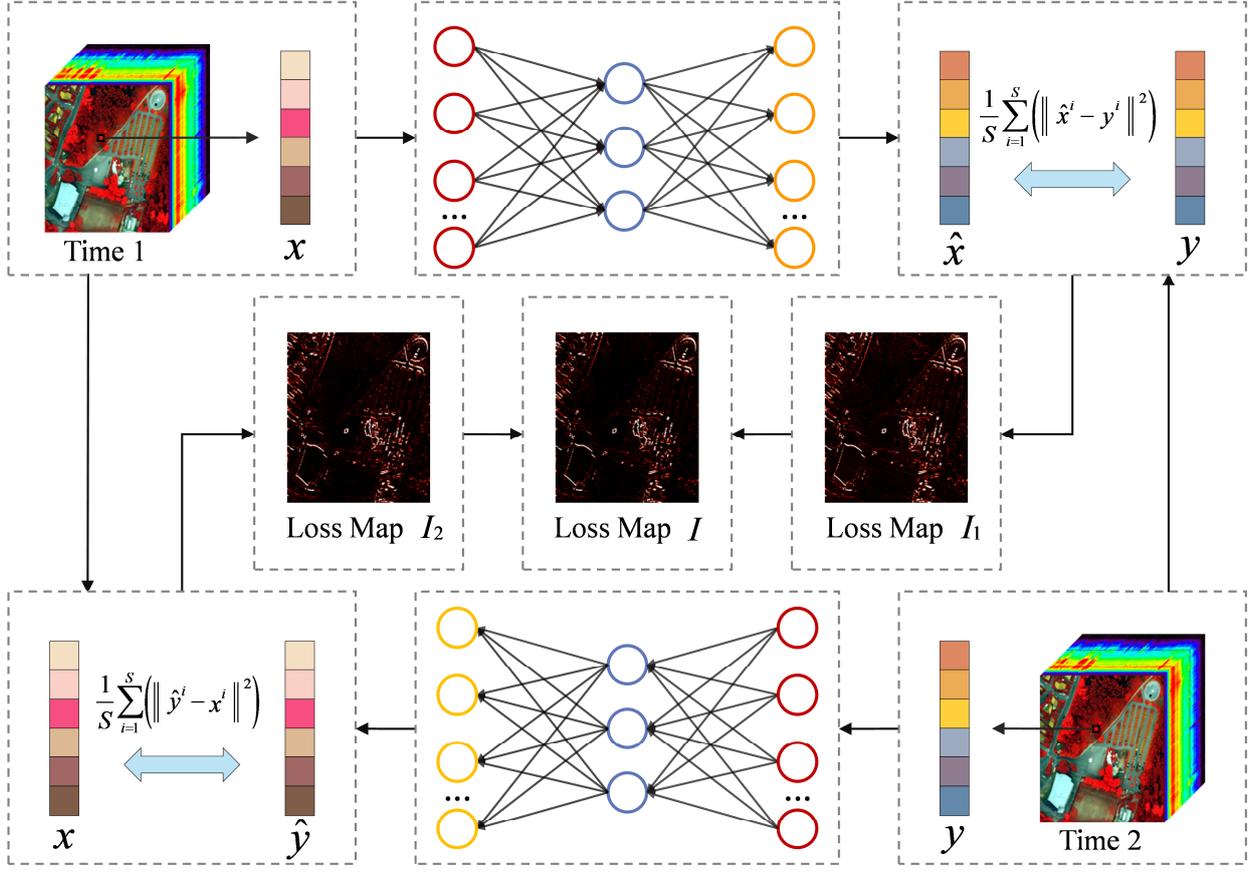

**Fig. 1.** The architecture of proposed ACDA. The top part of the network trains a predictor from HSI of Time 1 to HSI of Time 2, and the bottom part of the network trains a predictor from the HSI of Time 2 to Time 1. Then the predicted vector of Time 2, $\hat{x}$ and the original spectral vector of Time 2, $y$ is used to compute the loss map $I_1$. So is the loss map $I_2$. After two loss maps are acquired from two directions, the minimum of the two loss maps are computed as the final loss map $I$.

training model from $Y$ to $X$ can be described as:

$$L(\theta_2) = \frac{1}{S}\sum_{i=1}^{S}\left(\left\| \hat{y}^i - x^i \right\|^2\right) + \lambda_2 \sum_{j=1}^{n+1}\left\| w_j^2 \right\|^2 \quad (9)$$

where $\theta_2$ is the training parameter $\{w_1^2, b_1^2, \ldots, w_n^2, b_n^2, w_{n+1}^2, b_{n+1}^2\}$ of the second predictor. $w_j^2$ refers to weight parameter of different layer and $\lambda_2$ is the regularization coefficient. The loss map $I_2$ can be described as the MSE of spectral vector between $\hat{Y}$ and $X$ as follows:

$$I_2 = \mathrm{MSE}\left(\hat{Y}, X\right) \quad (10)$$

2) *Post-processing*

After two loss maps are obtained, we adopt the minimum operation on the two loss maps $I_1$ and $I_2$ in order to get better detection performance with less noise. Since the anomaly change detector adopts MSE, both loss maps cover bidirectional changes, such as appearance and disappearance of small objects. The final anomaly change intensity map $I$ can be expressed as follows:

$$I = \min(I_1, I_2) \quad (11)$$

The reasons of using min operation can be summarized as follows. 1) If a pixel belongs to anomalous change, the anomaly intensity value in both loss maps is large, and then the smaller one is also a big value. As a result, the anomalous information is preserved after min operation. 2) If a pixel does not have anomalous change, both of the intensity values in two loss maps are small, the minimum of those two makes it less possible to be an anomalous one. 3) If one of the two intensity values is large resulting from the training error or local environment, the minimum operation can cut down the possibility of being anomaly change, leading to less noise of the final result. Therefore, the min operation can absorb and amplify the advantage of the two loss maps.

3) *The scheme of proposed ACDA*

The detailed implementation of ACDA is depicted in **Algorithm 1**.

## III. EXPERIMENTAL RESULTS AND ANALYSIS

In order to illustrate the effectiveness of proposed method, abundant experiments on two benchmark hyperspectral datasets for HACD have been conducted. In this section, the description about the two real hyperspectral datasets and the experiments are firstly introduced in detail. Then the anomaly change detection performance and analyses are represented, which are followed by the parameter analyses about the hidden units and the selection of training samples for ACDA. Finally,

**Algorithm 1** Process of Training and Generating Anomaly Change Intensity Map for ACDA
__________________________________________________
**Input:**
Hyperspectral images $X$ and $Y$;
**Output:**
The anomaly change intensity map $I$;
1: Employ USFA pre-detection to generate training samples $x^i, i=1,2,...,S$ and $y^i, i=1,2,...,S$;
2: Initialize the AE network's parameters $\{\theta_1, \theta_2\}$;
3: **while** epoch < max_epochs do
4:  Calculate the predictive spectral vector $\hat{x}$ and $\hat{y}$;
5:  Calculate the loss function value $L(\theta_1)$ and $L(\theta_2)$;
6:  Back propagation to update the parameters with gradient descent algorithm;
7:  epoch++;
8: **end while**
9: Calculate the predicted image $\hat{X}$ and $\hat{Y}$;
10: Calculate the loss map $I_1 = \text{MSE}(\hat{X}, Y)$ and $I_2 = \text{MSE}(\hat{Y}, X)$;
11: Compute the final anomaly intensity map $I = \min(I_1, I_2)$;
12: **return** $I$;
__________________________________________________

we present the time cost analysis of four involved deep learning based algorithms.

*A. Hyperspectral datasets*

The benchmark datasets, "Viareggio 2013" datasets [39], include three hyperspectral images acquired at May 8-9, 2013 in Viareggio, Italy by an airborne hyperspectral sensor SIM.GA. The sensor collects spectral information ranging 400 nm - 1000 nm at a spectral resolution of 1.2 nm approximately. Besides, the ground spatial resolution is 0.6 m. As showed in **Fig. 2**, (a) is D1F12H1 which acquired at May 8, 2013, 14.18 (Greenwich Mean Time, GMT), (b) is D1F12H2 which acquired at May 8, 2013, 14.33 (GMT), and (c) is D2F22H2 which acquired at May 9, 2013, 12.64 (GMT). Since D1F12H1 and D1F12H2 are acquired at a cloudy day and the radiation comes from the scattered light of the sun, the imaging conditions of D1F12H1 and D1F12H2 are closely similar. While D2F22H2 is obtained on a clear sunny day, it is obvious that there is shadow in the region A, which indicates that D2F22H2 is under a space variant condition. In this paper, two groups of experiments are carried out, where D1F12H1 as well as D1F12H2 make up EX-1 while EX-2 consists of D1F12H1 and D2F22H2. These public datasets are available with five different preprocessing levels. We choose preprocessed data with de-striping, noise-whitening and spectrally binning. Concretely, the data was firstly processed by multi-linear regression to eliminate residual striping noise. And noise-whitening was used to normalize the noise variance. A spectral binning was also employed to reduce random noise by averaging the four consecutive spectral channels. For each HSI, there are 127 spectral bands with a scene size of 450×375.

It should be noted that all anomaly changes including appearance and disappearance are combined into the reference in spite of direction.

*B. Experimental setting*

In our ACDA model, there are three hidden layers with ReLU [23] as activation function and all weight along with bias matrices are initialized by he-normal way [40]. Using Adam [41] as optimizer, the ACDA are implemented by PyTorch. Besides, adding L2 regularization to loss function is good for avoiding over-fitting. For each experiment, regularization coefficient is selected as 0.001 by trial from the range [10e-8, 10e-1]. The training epochs are set as 200 in all experiments, and the learning rate batch size is 0.001 and 256, separately. As for the number of different hidden units, we design an AE network with three hidden layers, which are defined as $h_1$, $h_2$ and $h_3$, respectively. We denote ACDA-$h_1$-$h_2$ as ACDA model with the unit size of $h_1$ in the first hidden layer and $h_2$ in the second hidden layer. In the experiments, we adopt ACDA-60-40 model and the detailed analysis about the influence of the number of hidden units on algorithm performance is discussed in Section D.

Training samples are selected from pre-detected results of USFA instead of manual selection. For each experiment, 10000 samples, which are 6% approximately of the total number of the pixels, are chosen from the pre-detected results at random. Specifically, the number of bands whose 1/eigenvalue is greater than 1 is automatically selected for the anomaly detection of USFA. Then the anomaly intensity result is sent to K-means classifier with 3 clusters and the cluster samples with the smallest cluster center are assigned as the desired training set. Owning to the random initialization of weight matrices and bias vectors, we take the average of ten independently repeated results as the final anomaly intensity map. And the assessment is based on the average result.

Ten comparable algorithms are conducted on the datasets mentioned above, including Difference Reed-Xiaoli algorithm (Diff-RX) [22], Straight Anomalous Change Detector (SACD), Simple Difference Anomalous Change Detector (SDACD), Simple Difference Hyperbolic Anomalous Change Detector(SDHACD) [42], Chronochrome (CC) [17], Covariance Equalization (CE) [18], three fully connected (FC) [43] neural networks and Unsupervised Slow Feature Analysis between probability density functions of HSIs. CC and CE are two canonical predictor methods based on the statistical features of the images. As for the three fully connected neural networks, they are all predictor models for unusual change detection, which uses no labeled data and a three-layer quick prop-trained neural network to build a nonlinear predictor used for multi-spectral and panchromatic data sets. We refer to this algorithm here for HACD. The training settings and cost function adopted by the FC networks are the same as that of ACDA, but the number of hidden layer units is greater than or equal to the number of input layer units. Concretely, each network has three hidden layers and refers to FC-$h_1$-$h_2$, with the first hidden layer having the same number of cells as the third. The three networks, FC-127-127, FC-127-200 and FC-200-200, are specifically designed to be compared with bottleneck structure in this paper. And USFA is the pre-detection algorithm used in this paper which tries to extract the invariant features and detects the anomalies from transformed difference features. Various types of predictor models are used for comparison, including CC, CE, FC-127-127, FC-127-200 and FC-200-200. Note that the min operation is also used for these predictor models to produce the final results for them. Besides, all the methods based on deep learning are repeated ten times and average of the anomaly intensity maps is taken as the final anomaly change map as well as the evaluation.

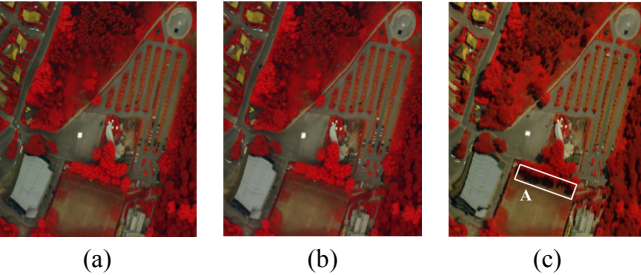

(a)          (b)          (c)

**Fig. 2.** The Viareggio datasets used for experiments. (a) D1F12H1, (b) D1F12H2, (c) D2F22H2.

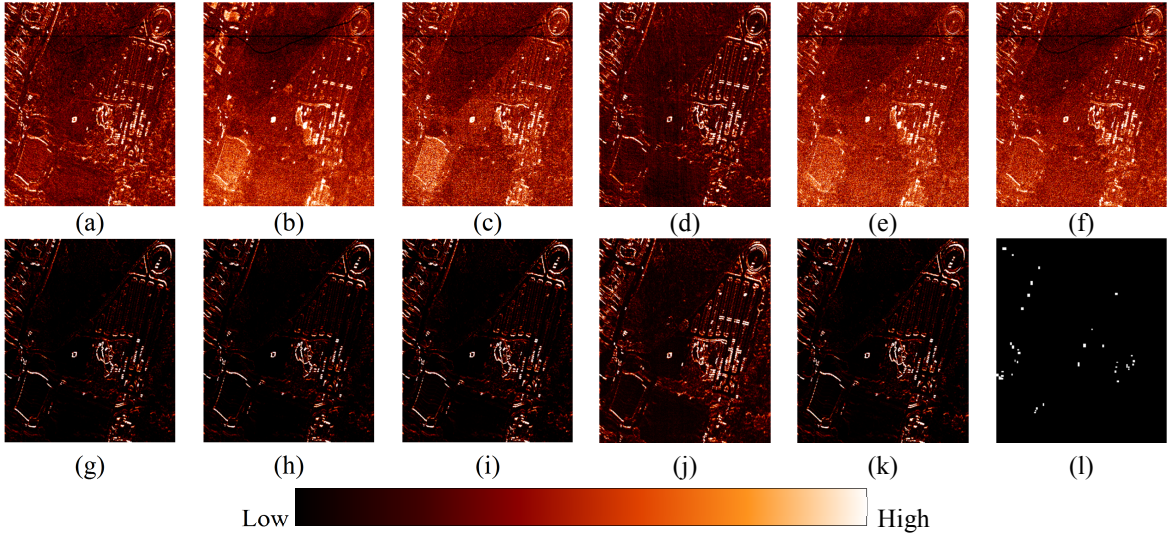

**Fig. 3.** The anomaly change intensity map of D1F12H1-D1F12H2. (a) Diff-RX, (b) SACD, (c) SDACD, (d) SDHACD, (e) CC, (f) CE, (g) FC-127-127, (h) FC-127-200, (i) FC-200-200, (j) USFA, (k) ACDA-60-40, (l) Ground truth map.

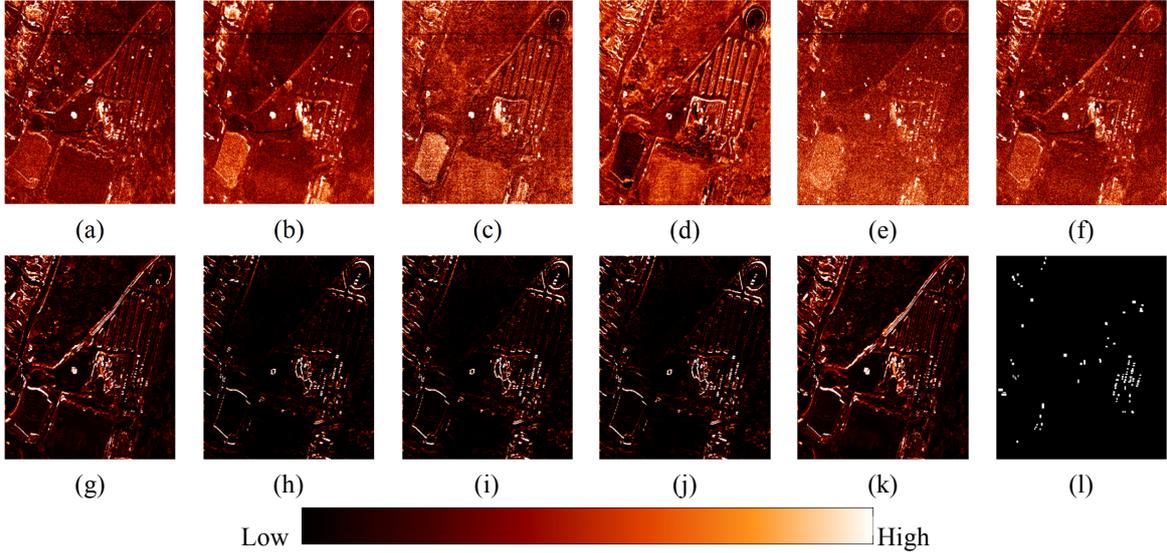

**Fig. 4.** The anomaly change intensity map of D1F12H1-D2F22H2. (a) Diff-RX, (b) SACD, (c) SDACD, (d) SDHACD, (e) CC, (f) CE, (g) FC-127-127, (h) FC-127-200, (i) FC-200-200, (j) USFA, (k) ACDA-60-40 (l) Ground truth map.

Moreover, Receiver Operating Characteristic (ROC) [44] curve as well as Area under Curve (AUC) are adopted for quantitative assessments. ROC is widely used for no threshold to test the performance of anomalous change detection. The horizontal axis of the ROC curve is the false alarm rate and the vertical axis is the detection rate. In details, The closer the ROC is to the upper left corner, the better the performance of the method. With the same false alarm rate, the method with higher detection rate gains superior detection result. And the AUC is the area enclosed by ROC curve, X and Y axis. The bigger the AUC is, the greater the algorithm is.

*C. Results and analysis*

In this subsection, the detection results of the experiments are firstly analyzed. Then the analysis of the impact of the min operation on the final results is presented.

**Fig. 3** and **Fig. 4** show the anomaly change intensity maps of D1F12H1-D1F12H2 and D1F12H1-D2F22H2 by Diff-RX, SACD, SDACD, SDHCD, CC, CE, FC-127-127, FC-127-200, FC-200-200, USFA, and ACDA-60-40, respectively. There are all 2% linearly stretched. For the D1F12H1-D2F22H2 dataset, there are 849 pixels of anomaly changes showed in **Fig. 3**(l). And **Fig. 4**(l) is the ground truth maps of D1F12H1-D2F22H2, with 1778 pixels of anomalous changes. The brighter region is more likely to be the anomaly change. As show in **Fig. 3**, most of the bright region in the result of SACD, SDHACD, FC-200-200 and ACDA-60-40 correspond to the anomalous changes of the ground truth map for D1F12H1-D1F12H2 dataset. And there are lots of noises on results of Diff-RX, SACD, SDACD, CC and CE. By contrast, FC-127-127, FC-127-200, FC-200-200, USFA, and ACDA-60-40 are effective in suppressing the background, where the methods based on deep learning work better. For D1F12H1-D2F22H2 dataset, it is obvious that SACD, CE, USFA and ACDA-60-40 detect most of anomaly changes presented on **Fig. 4.** Yet there are lots of noises and pseudo changes in the result of Diff-RX, SACD, SDACD, CC and CE, which are consistent with the results of D1F12H1-D1F12H2. Since D2F22H2 are acquired on a sunny day and there are large distinct shadows of the vegetation on the image, it is a challenge to deal with huge gaps of the imaging conditions. Except for several algorithms seriously affected by noise mentioned above, e.g., SACD, the analyses about the performance of the other algorithms on the shadow problem are summarized as follows. It is apparent that there are high anomaly intensity values in the shadow area for the results of SDHACD and USFA. As for SDHACD, it belongs to the space invariant assumption and supposes that the spectral vector is sampled from the multivariate Gaussian model. But D1F12H1-D2F22H2 dataset is under a variant background space and the dataset might not obey the distribution perfectly. And USFA is designed to suppress the spectral differences of the slowly varying pixels to highlight the changed ones. But the performance of USFA in this case is not satisfying, which may lie in the large differences of the shadow area hard to be depressed. Whereas the ACDA-60-40, FC-127-127, FC-127-200 and FC-200-200 suppress the big spectral differences greatly, of which ACDA-60-40 detects more anomaly changes. It is indicted that ACDA-60-40 shows good performance on shadow suppression and anomaly change detection. The powerful nonlinearity of neural networks promotes the effect of the predictor methods based on deep learning. Furthermore, compared to the other FC networks, the special structure bottleneck enables AE to extract the essential features of the input and is critical to establish a valid mapping relationship between the two backgrounds.

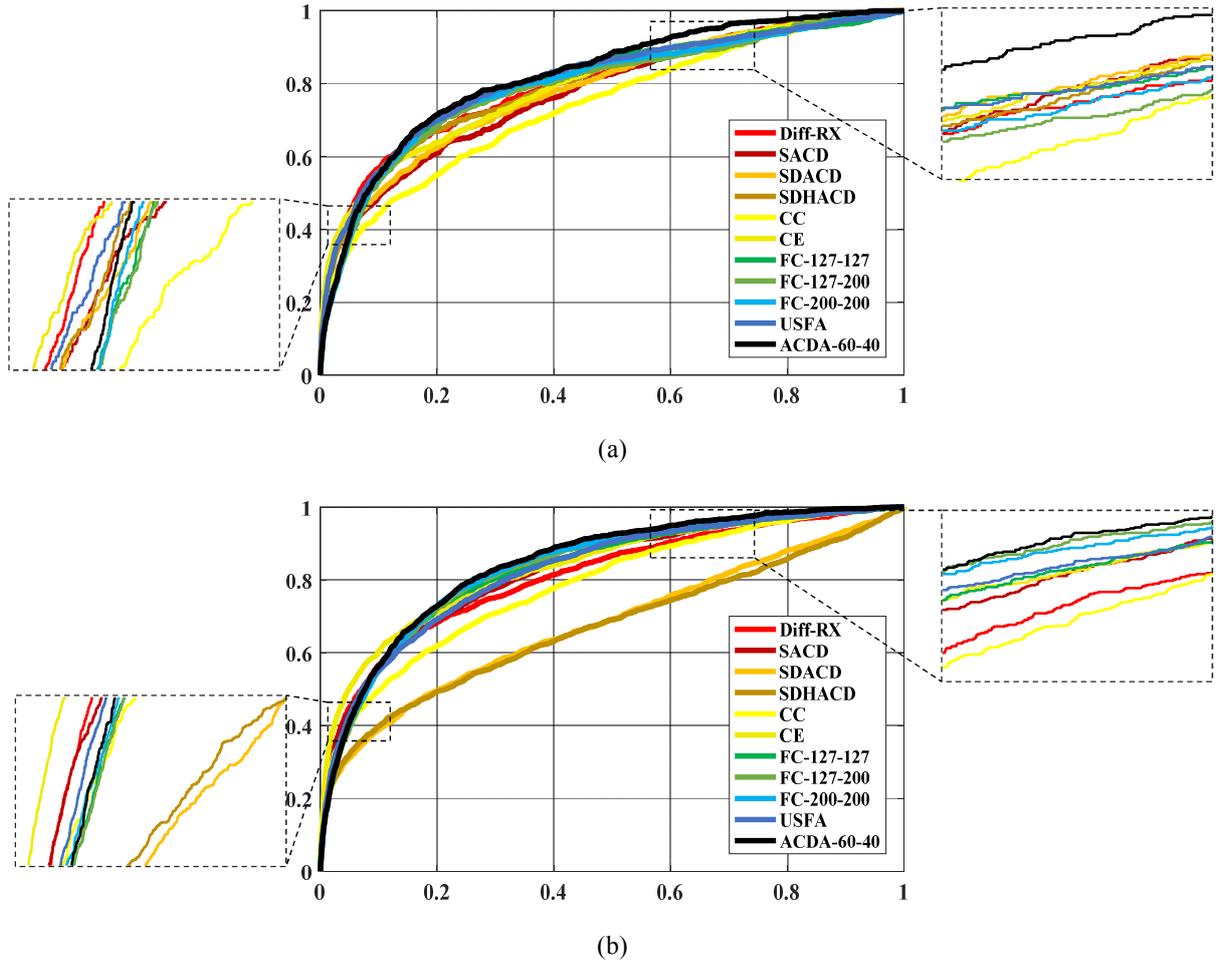

**Fig. 5.** The ROC performance for two data sets. (a) D1F12H1-D1F12H2, (b) D1F12H1-D2F22H2.

**Fig. 5** represents the ROC performance of the eleven techniques for two data sets and the details are zoomed in. It can be observed that the ROC of ACDA-60-40 in black is closest to the top left. Concretely, when the false alarm rate is under the low level range, CE, Diff-RX, and USFA all show high detection rate from the left enlarged view of **Fig. 5**(a) and (b). And ACDA-60-40 obtains best detection rate from the right enlarged view under high false alarm rate range. **Fig. 6** presents the quantitative evaluation of two experiments. The bigger AUC value refers to greater effect. And the maximum is highlighted in bold in figures. For D1F12H1-D1F12H2, the proposed method has the largest AUC values of 0.8221. The performance of USFA ranks second with 0.8112, followed by FC-127-127 with 0.8019, FC-200-200 with 0.7996 and Diff-RX with 0.7984. And other approaches based on deep learning also perform well, especially compared with SACD and CC. In addition, CC works worst at an AUC value of 0.7412, which may be attributed to the impact of mis-registration. In the case of D1F12H1-D2F22H2, the proposed method gains the largest AUC values again with 0.8451. FC-200-200, FC-127-200 and CE also acquire satisfying results with 0.8387, 0.8365 and 0.8359. Besides, the AUC values of SDACD and SDHACD is extremely low at 0.6778 and 0.6708, respectively, whereas SACD gains a nice AUC value with 0.8252. These three methods are all under the multivariate Gaussian model, but the performances are quite different. The reason may be that the SACD adopts joint vector observation model while SDACD and SDHACD both employ the difference vector observation model, since the spectral differences between the two images are quite large. The quantitative assessments of two experiments manifest that the proposed ACDA has best performance on anomaly change detection and background suppression.

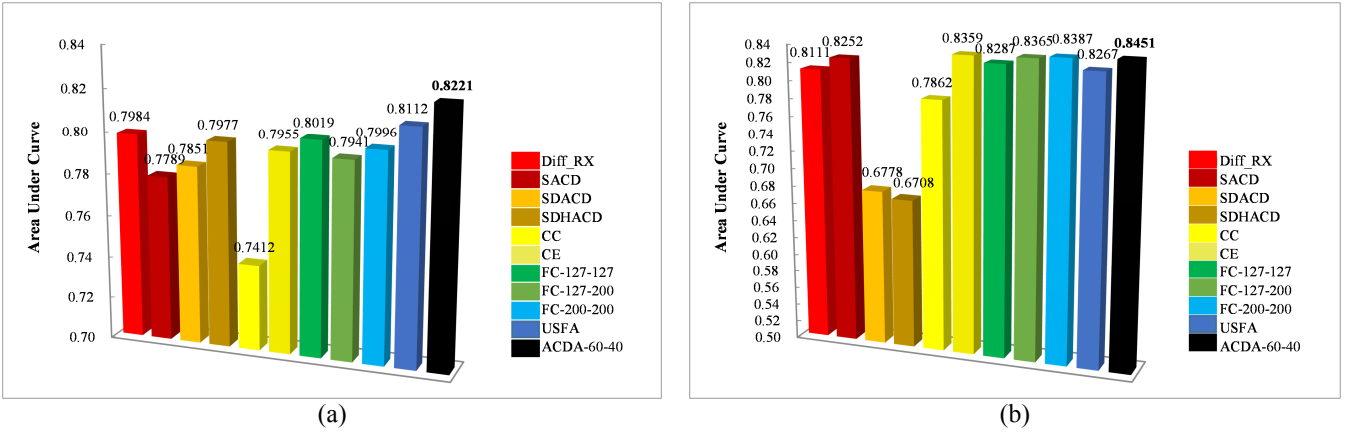

Fig. 6. The ROC performance for two data sets. (a) D1F12H1-D1F12H2, (b) D1F12H1-D2F22H2

And there is another trick need to be discussed as follows. For each experiment, there are two predictors trained to reap two loss maps. The final result comes from the min operation of the two loss maps as above mentioned. **Fig. 7** displays the influence of min operation on the final loss maps about the first experiment. As **Fig. 7** shows, it can be known from the ground truth map that region B has anomaly changes, which is also bright in the loss map $I_1$ and $I_2$. And the min operation still reserves the bright anomalous information. For region C, there is no anomalous change in fact. But there is still some twinkling spots detected in region C of loss map $I_1$. And region C in loss map $I_2$ seems darker. The min operation takes the minimum of the two anomaly intensity values as the final results, leading to smaller anomaly values and reducing the probability of being anomaly changes and less noise. Fig.8 shows another example on D1F12H1-D2F22H2. Region A is the shadow area and does not have anomaly changes. But there are large spectral differences between two HSIs. As **Fig. 8** shows, region A in loss map $I_1$ is bright and the whole shadow area is likely to be detected as anomalous change. While region A in loss map $I_2$ has low anomaly intensity value and is less possible to be anomalous change. So the minimum result of the two loss maps yields low possibility of being anomalous change. And when a region such as region B has high values in both two loss maps, the result of the min operation can still preserve the high intensity information.

All in all, when a pixel has anomaly change in fact, the anomaly intensity values detected by two predictors are both high. The minimum of them is still a large value which still retains anomaly change information. When a pixel does not contain anomaly change actually, the anomaly intensity values detected by two predictors are both low. The minimum of them becomes lower which makes the pixel less likely to be anomalous change. Another alternative max operation is proved to be less effective than min operation by trial. Taking the region A for example, the maximum of the anomaly value on two loss maps is the one on loss map $I_1$, which makes it high likely to be anomaly change. The min operation reduces the noise on the final intensity map but at the expense of some anomaly changes, which brings about low detection rate under low false alarm rate range.

### D. Parameter analysis

This subsection discusses the influence of the units of hidden layers and the selection of training samples on the experiment performance.

The number of the units of hidden layers is a group of crucial parameters of ACDA. Considering that deeper networks aggravate the time cost and the possibility of over-fitting, we design an ACDA model based on auto-encoder with three

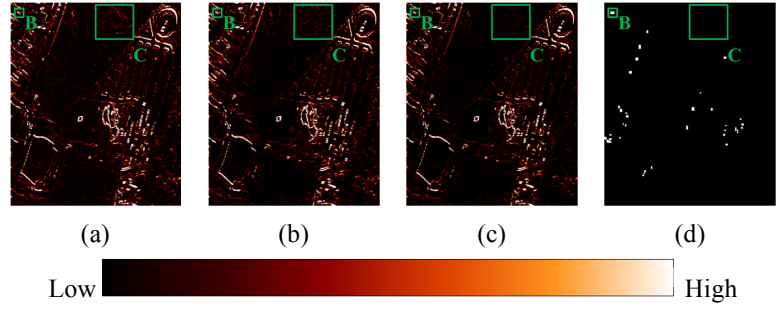

(a) (b) (c) (d)

Low　　　　　　　　　　　　　　　　High

**Fig. 7.** The influence of min operation on the final loss maps about the D1F12H1-D1F12H2. (a) Loss map $I_1$ of predictor $X \to Y$, (b) Loss map $I_2$ of predictor $Y \to X$, (c) Final loss map $I$, (d) Ground truth map.

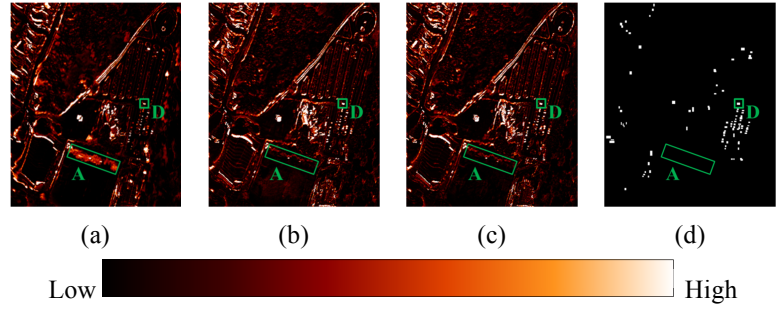

(a) (b) (c) (d)

Low　　　　　　　　　　　　　　　　High

**Fig. 8.** The influence of min operation on the final loss maps about the D1F12H1-D2F22H2. (a) Loss map $I_1$ of predictor $X \to Y$, (b) Loss map $I_2$ of predictor $Y \to X$, (c) Final loss map $I$, (d) Ground truth map.

**Table 1**

AUC comparison under different parameter $h_1$ and $h_2$ for D1F12H1-D1F12H2

| $h_2$ | $h_1$ | | | | |
|---|---|---|---|---|---|
| | 120 | 100 | 80 | 60 | 40 |
| 100 | 0.8021 | —— | —— | —— | —— |
| 80 | 0.8225 | 0.8133 | —— | —— | —— |
| 60 | 0.8035 | 0.7959 | 0.8089 | —— | —— |
| 40 | 0.8102 | 0.7940 | 0.8112 | 0.8221 | —— |
| 20 | 0.8048 | 0.7981 | 0.8004 | 0.8157 | 0.8159 |

**Table 2**

AUC comparison under different parameter $h_1$ and $h_2$ for D1F12H1-D2F22H2

| $h_2$ | $h_1$ | | | | |
|---|---|---|---|---|---|
| | 120 | 100 | 80 | 60 | 40 |
| 100 | 0.8402 | —— | —— | —— | —— |
| 80 | 0.8374 | 0.8443 | —— | —— | —— |
| 60 | 0.8334 | 0.8399 | 0.8379 | —— | —— |
| 40 | 0.8417 | 0.8364 | 0.8358 | 0.8451 | —— |
| 20 | 0.8442 | 0.8451 | 0.8428 | 0.8451 | 0.8359 |

hidden layers. Since the configuration of neural network can be various and there is no accurate means to choose the optimum number of the units of hidden layers, we set the range of $h_1$ as [120 100 80 60 40] and $h_2$ as [100 80 60 40 20] to tune the parameters. The size of the input layer is equivalent to the channel of the HSI, which equals to 127. For the special structure bottleneck of AE mode, $h_1$ starts from 120 and $h_2$ starts from 100. **Table 1** and **Table 2** is the evaluation of AUC results on D1F12H1-D1F12H2 and D1F12H1-D2F22H2. According to **Table 1** and **Table 2**, the AUC decreases when the number of $h_1$ is very large. This phenomenon can be explained as follows. When the number of $h_1$ and $h_2$ are both large, the code feature with high dimension cannot represent the essence of the input. When the number of $h_1$ is large and $h_2$ is small, the gaps between these two dimensions lead to a crack in the process of feature transformation and flow, thus losing part of features. In addition, when the number of $h_1$ is too small, the AUC turns low, which can be attributed to the large dimensional difference between the two connecting layers. Thus we select $h_1$ =60 and $h_2$ =40 as the optimal result for both experiments.

**Fig. 9** represents the impact of three different ways of training samples selection on the effectiveness of experiments. Random strategy and USFA refer to choose samples from the whole images and the pre-detection results of USFA at random, respectively. And the ground truth strategy refers to select the samples from the ground truth map. As **Fig. 9** shows, the results of both experiments show the same pattern, which can summarized as two points. Firstly, the results trained by USFA and ground truth are very similar, which demonstrates the effectiveness of the USFA. Secondly, there is slight difference between the AUC performance of random selection and USFA selection. This is because there are a small number of anomaly changes in the ground truth map. And the total number of samples used for training only occupies around 6% of the total number of pixels. Though the probability is very small that the pixels of anomalous change are selected, it cannot be ruled out. And once pixels of anomaly change are fed into the model, mapping relationship between different imaging conditions would be contaminated, thus impairing the effect of predictive model.

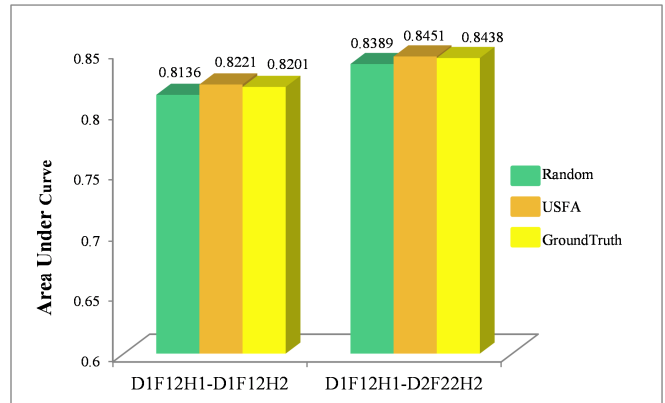

**Fig. 9** The influence of different training sample selection strategies on the AUC performance of two data sets.

**Table 3**
Running time of different deep learning approaches

| Dataset | FC-127-127 | FC-127-200 | FC-200-200 | ACDA-60-40 |
| --- | --- | --- | --- | --- |
| D1F12H1-D1F12H2 | 157.92 ± 5.21(s) | 159.15 ± 4.20(s) | 159.59 ± 2.98(s) | **152.29 ± 4.43(s)** |
| D1F12H1-D2F22H2 | 160.42 ± 4.81(s) | 159.22 ± 5.77(s) | 153.56 ± 4.07(s) | **137.23 ± 4.66(s)** |

*E. Run time cost analysis*

The time cost of the method is significant for timely detection in practical application. There are four algorithms based on neural networks involved in this paper, which are FC-127-127, FC-127-200, FC-200-200, and ACDA-60-40. **Table 3** displays the run time analysis of them. And the CPU used in the paper is Intel Xeon E3-1220 3.00-GHz processor with RAM as 16 GB. And the GPU adopts a NVIDIA RTX 2080 TI graphic card. According to **Table 3**, the proposed method runs the fastest in both experiments owing to smaller size of network. The time consumption of the other approaches is a little longer but within acceptable range.

## IV. CONCLUSION

In this paper, we point out that the classical linear predictors have limited effect in dealing with the problem of variant space. And a nonlinear predictive model based on auto-encoder has been proposed for hyperspectral anomaly change detection, which maps a hyperspectral image to the imaging condition of another and get a predictive image. The bottleneck structure of AE is capable of extracting essence feature from input. Rather than replicating the output from the input, we design the training out as the corresponding spectral vector of another multi-temporal HIS to construct predictor model. And the min operation between the loss maps helps to reduce the noise of the final result. In the experiments, the performances on public hyperspectral data sets demonstrate that the proposed ACDA outperforms than other state-of-the-art techniques, especially compared with another five classical and deep learning based predictor approaches. On the whole, the discussion can be summarized as follows.

1) The designed bottleneck structure is proved to perform better than normal FC network which uses the largest size of unit at hidden layer.

2) The ACDA model performs well under the imaging condition of space variability, which is a challenging problem of HACD.

3) As for how to build an appropriate architecture for practical application, the results of parameter analysis indicate that the gaps of the number of units between adjacent hidden layers should be uniformly decreasing for encoder, and uniformly increasing for decoder.

4) The ACDA runs the fastest among all deep learning based methods.

Since the ACDA mainly focuses on the spectral features of multi-temporal HSIs, we will put more attention on the combination of spectral and spatial features on further study for complex space variant background changes.